\documentclass[11pt]{article}
\usepackage{latexsym}  
\usepackage{amssymb}
\usepackage{graphicx}
\usepackage{amsmath}

\newcommand{\be}{\begin{equation}}
\newcommand{\ee}{\end{equation}}
\newcommand{\bea}{\begin{eqnarray}}
\newcommand{\eea}{\end{eqnarray}}
\newcommand{\bref}[1]{(\ref{#1})}
\begin{document}
\begin{titlepage}
\begin{flushright}
\today
\end{flushright}
\vspace{4\baselineskip}
\begin{center}
{\Large\bf  Universal Mass Matrix and Leptonic $\theta_{13}$ Angle.}
\end{center}
\vspace{1cm}
\begin{center}
{\large Hiroyuki Nishiura$^{a,}$
\footnote{E-mail:nishiura@is.oit.ac.jp}}
and
{\large Takeshi Fukuyama$^{b,c,}$
\footnote{E-mail:fukuyama@se.ritsumei.ac.jp}}
\end{center}
\vspace{0.2cm}
\begin{center}
${}^{a} $ {\small \it Faculty of Information Science and Technology, 
Osaka Institute of Technology,\\ Hirakata, Osaka 573-0196, Japan}\\[.2cm]

${}^{b}$ {\small \it Department of Physics and R-GIRO, Ritsumeikan University,
Kusatsu, Shiga,525-8577, Japan} \\and\\
${}^{c}$ {\small \it Maskawa Institute for Science and Culture,
Kyoto Sangyo University, Kyoto 603-8555, Japan}
\medskip
\vskip 10mm
\end{center}
\vskip 10mm
\begin{abstract}
We propose a universal mixing hypothesis between quark and lepton sectors 
at high energy scale (probably GUT scale) where quark-lepton universality holds.  
Namely in the charged lepton diagonal base, all the other mass matrices for up and down quarks and neutrinos 
are diagonalized by the same unitary matrix except for the phase elements. 
Thanks to this hypothesis, the observed values of the Cabibbo-Kobayashi-Maskawa (CKM) quark mixing matrix and the mixing angles 
$\theta_{12}$ and $\theta_{23}$ in the Maki-Nakagawa-Sakata (MNS) lepton mixing matrix can predict the unknown magnitudes 
of the mixing angle $\theta_{13}$  and of the CP violating Dirac phase $\delta $ 
in the MNS matrix. 
Their allowed regions are rather narrow, 
$0.036<\mbox{sin}\theta_{13}<0.048$ and $6^\circ<\delta  <12^\circ$.
\end{abstract}
\end{titlepage}

Over thirteen years before we proposed first in the world $\mu-\tau$ symmetry in the neutrino mass matrix model \cite{F-N}, 
\begin{equation}
M_{light}=
\left(
\begin{array}{ccc}
0&A&A\\
A&B&C\\
A&C&B
\end{array}
\right).
\label{form}
\end{equation}
Since neutrino oscillation experiment is wholly insensitive to the Majorana phases, 
the MNS lepton mixing matrix is in general written in the form
\begin{equation}
U=
\left(
\begin{array}{ccc}
c_{13}c_{12},&c_{13}s_{12},& s_{13}e^{i\delta}\\
-c_{23}s_{12}+s_{23}c_{12}s_{13}e^{-i\delta},&c_{23}c_{12}+s_{23}s_{12}s_{13}e^{-i\delta},&-s_{23}c_{13}\\
-s_{23}s_{12}-c_{23}c_{12}s_{13}e^{-i\delta},&s_{23}c_{12}-c_{23}s_{12}s_{13}e^{-i\delta},&c_{23}c_{13}
\end{array}
\right) .
\label{mixing}
\end{equation}
Here, $c_{ij}=\mbox{cos}\theta_{ij},~s_{ij}=\mbox{sin}\theta_{ij}$ as usual.
If we adopt $\theta_{23}=\pi/4$ and $\theta_{13}=0$, then the neutrino mass matrix becomes
\begin{eqnarray}
M_{light}&=&U
\left(
\begin{array}{ccc}
-m_1&0&0\\
0&m_2&0\\
0&0&m_3
\end{array}
\right)
U^T\nonumber\\
&=&\left(
\begin{array}{ccc}
-c_{12}^2m_1+s_{12}^2m_2&{1\over\sqrt{2}}c_{12}s_{12}(m_1+m_2)&{1\over\sqrt{2}}c_{12}s_{12}(m_1+m_2)\\
{1\over\sqrt{2}}c_{12}s_{12}(m_1+m_2)&{1\over 2}(-s_{12}^2m_1+c_{12}^2m_2+m_3)&-{1\over 
2}(s_{12}^2m_1-c_{12}^2m_2+m_3)\\
{1\over\sqrt{2}}c_{12}s_{12}(m_1+m_2)&-{1\over 2}(s_{12}^2m_1-c_{12}^2m_2+m_3)&{1\over 
2}(-s_{12}^2m_1+c_{12}^2m_2+m_3)
\end{array}
\right).\nonumber\\
\label{mixing2}
\end{eqnarray}
In \bref{mixing2} if we further assume 
\be
-c_{12}^2m_1+s_{12}^2m_2=0,
\label{small}
\ee
then we obtain the mass matrix of \bref{form}. The assumption given 
in \bref{small} favored the small mixing angle solution which was still surviving at the time when \cite{F-N} was written.
Also the vanishing (1,1) component is interested in connection with seesaw invariant mass matrix \cite{Fuku2}.
However, KamLAND \cite{KamLAND} selected the larger part of solar neutrino angles, 
and we may relax the vanishing condition of (1,1) component.
In this paper we simply accept the neutrino oscillation data\cite{KamLAND2, MINOS}, 
\be
\mbox{tan}^2\theta_{12}=0.47_{-0.05}^{+0.06},~~ \mbox{sin}^2 2\theta_{23}=1
\label{oscillation data}
\ee
in \bref{mixing} and we denote the lepton mixing matrix $U$ with fixing $\theta_{12}$ and $\theta_{23}$ 
by \bref{oscillation data} as $U_{SA}$.  
Then unknown parameters in the lepton mixing matrix are only $\theta_{13}$ 
and the Dirac CP violating phase $\delta$. 
Namely, in this paper,  we accept the deviation from tribimaximal \cite{TBM} and $\mu-\tau$ symmetric models.
In \cite{Koide} we generalized $\mu-\tau$ symmetry of \bref{form} 
to quark (and heavy neutrino) mass matrices (Universal texture model with $2-3$ 
symmetry).
In the present paper we consider the universal character of not \bref{form} but \bref{mixing} 
in the quark and lepton sectors. 
That is, we assume that mass matrices for up and down quarks are diagonalized 
by the same mixing matrix $U_{SA}$ ($U_d=U_u=U_{SA}$) in the limit of some flavor symmetry(Universal hypothesis), and the breaking of the symmetry comes into existence  though diagonal phase matrices $P_d$ and $P_u$.
In this model, the mass matrix for up (down) quarks is diagonalized by a following mixing matrix $U_u$ ($U_d$).
\be
U_d=P_d^\dagger U_{SA},~~U_u=P_u^\dagger U_{SA}.
\label{proposition}
\ee
So the CKM quark mixing matrix is represented by
\be
V_{CKM}=U_{SA}^\dagger P_uP_d^\dagger U_{SA}\equiv U_{SA}^\dagger P U_{SA},
\label{CKM}
\ee
where
\be
P\equiv P_uP_d^\dagger=\left(
\begin{array}{ccc}
e^{i\phi_1}&0&0\\
0&e^{i\phi_2}&0\\
0&0&1\\
\end{array}
\right).
\ee

If neutrino is Majorana particle, the MNS matrix has additional two Majorana phases. 
However, it does not affect the subsequent discussions. 
The universality implies that quark-lepton is considered as a multiplet, for instance, like ${\bf 16}$ in SO(10). 
So we cannot use rebasings independently in both quark and lepton sectors. 
The subsequent discussions are considered in the base where the mass matrix for the charged leptons is diagonal.
The $V_{CKM}$ has four physical parameters which are base independent, although $\phi_1$ and $\phi_2$ are base dependent.  
The hypothesis asserts the existence of the base 
where (6) holds, which is far from being trivial. Important is that the physical 
parameters $\theta_{13}$ and $\delta$ obtained in this model are base independent.
If this is concerned with mass relation at GUT scale, we must consider renormalization group equation (RGE) effect on it.
The RGE effect for all Yukawa couplings of quarks and leptons as well as heavy Majorana neutrinos was discussed in \cite{F-O}\cite{Antusch}.
However, we do not consider RGE since it is rather small except for degenerate neutrino mass case \cite{Mohapatra}. 

\paragraph{Numerical fitting}@\\

We treat $\theta_{13}$ and $\delta$ as free parameters in $U_{SA}$. 
Therefore, the CKM quark mixing matrix elements are determined 
as functions of $\theta_{13}$, $\delta$, $\phi_1$, and $\phi_2$ from \bref{CKM}.
In the following numerical fitting, for each set of fixed values of $s_{13} ( = \sin\theta_{13})$ and $\delta$, 
we search for the solution of phase parameters $\phi_1$ and $\phi_2$ 
which give the observed values of the CKM elements. 
Fig. 1 shows an allowed region in the $\phi_1-\phi_2$ plane 
which is consistent with the experimental data of the CKM elements, 
$|(V_{CKM})_{us}|$, $|(V_{CKM})_{cb}|$, $|(V_{CKM})_{ub}|$, and $|(V_{CKM})_{td}|$ 
for a typical set of parameters $s_{13}=0.042$ and $\delta=9^\circ $.
Here, we use\cite{PDG} 
\begin{eqnarray}
|(V_{CKM})_{us}|&=&0.2255 \pm 0.0019,\\
|(V_{CKM})_{cb}|&=&0.0412 \pm 0.0011,\\
|(V_{CKM})_{ub}|&=&0.00393 \pm 0.00036,\\
|(V_{CKM})_{td}|&=&0.0081 \pm  0.0006.
\end{eqnarray}  
We search for all the possible sets of parameters $s_{13}$ and $\delta$ by using the same method as above typical case. 
Then we can derive allowed regions in the $\delta-s_{13}$ plane, 
in which we have solutions for $\phi_1$ and $\phi_2$ consistent with the experimental data of the CKM elements. 
In Fig. 2 , we present allowed regions in the $\delta-s_{13}$ plane 
which can reproduce the above four experimental data of the CKM elements. 
It shows the following allowed region for $s_{13}$ and $\delta$
\be
0.036<s_{13}<0.048
\ee
and 
\be
6^\circ <\delta<12^\circ .
\ee
In this parameter region, the hypothesis \bref{proposition} is consistent with the data for the CKM elements.  
Namely, the hypothesis predicts small value for $s_{13}$.
This value is small by order one compared with the recent analysis \cite{Fogli}, 
$
s_{13}^2=0.02\pm0.01~(1\sigma)
$.
This does not preclude our result since the error bar is still so large. 
The CP violating phase $\delta $ is predicted to have a magnitude around 10$^\circ$.

\paragraph{Mass matrix}@\\

So far, we have discussed numerical fittings model-independently.
Hereafter, we study the implication of the universal property adopted in the preceding arguments to the mass matrix. 
For simplicity, we replace the universal mixing matrix by a tribimaximal matrix \cite{TBM} up to the phase factor given by 
\be
U_{TB}(\delta )=\left(
\begin{array}{ccc}
c\sqrt{\frac{2}{3}},&c\sqrt{\frac{1}{3}},&se^{-i\delta}\\
-\sqrt{\frac{1}{6}}+s\sqrt{\frac{1}{3}}e^{i\delta},&\sqrt{\frac{1}{3}}+s\sqrt{\frac{1}{6}}e^{i\delta},&-c\sqrt{\frac{1}{2}}\\
-\sqrt{\frac{1}{6}}-s\sqrt{\frac{1}{3}}e^{i\delta},&\sqrt{\frac{1}{3}}-s\sqrt{\frac{1}{6}}e^{i\delta},&c\sqrt{\frac{1}{2}}\\
\end{array}
\right),
\ee
where we have abbreviated $s_{13}=s,~c_{13}=c$. 
Here $s_{12}=\frac{1}{\sqrt{3}}$ and $s_{23}=\frac{1}{\sqrt{2}}$ are assumed in $U$ for simplicity. 
As is well known, this is a good approximation and sufficient for a study of the structure of universal mass matrices.
The mass matrices $M_u$ and $M_d$ for up  and down quarks in our scheme are respectively given by 
\be
M_u=P_u^\dagger \hat{M}_uP_u,~~M_d=P_d^\dagger \hat{M}_dP_d,
\ee
where
\be
\hat{M}_u=U_{TB}(\delta)\hat{D}_uU_{TB}^\dagger(\delta),~~\hat{M}_d=U_{TB}(\delta)\hat{D}_dU_{TB}^\dagger(\delta).
\ee
Here, $\hat{D}_u$ and $\hat{D}_d$ are diagonalized mass matrices given by
\be
\hat{D}_u=\mbox{diag}(m_1^{(u)},m_2^{(u)},m_3^{(u)}),~~\hat{D}_d=\mbox{diag}(m_1^{(d)},m_2^{(d)},m_3^{(d)}).
\ee
It should be noted that the unitary matrix $U_{TB}(\delta)$ is common in up and down quarks (and neutrinos).
That is, 
\be
\hat{M}_i=U_{TB}(\delta)\left(
\begin{array}{ccc}
m_1^{(i)}&0&0\\
0&m_2^{(i)}&0\\
0&0&m_3^{(i)}
\end{array}
\right)U_{TB}^\dagger(\delta),
\ee
where i=up, down quarks, and light neutrinos.

Let us separate the mass matrix into two parts.
The first part is a main part with $s=0$ which is 2-3 symmetric, 
and the second part is small correction part due to $s\neq0$ which is 2-3 antisymmetric.  
For $s=0$, the mass matrix $\hat{M}$ is written in terms of mass eigen values as,
\be
\hat{M}(s=0)=\left(
\begin{array}{ccc}
\frac{1}{3}(m_2+2m_1)&\frac{1}{3}(m_2-m_1)&\frac{1}{3}(m_2-m_1) \\
\frac{1}{3}(m_2-m_1)&\frac{1}{6}(3m_3+2m_2+m_1)&\frac{1}{6}(-3m_3+2m_2+m_1)\\
\frac{1}{3}(m_2-m_1)&\frac{1}{6}(-3m_3+2m_2+m_1)&\frac{1}{6}(3m_3+2m_2+m_1)
\end{array}
\right).
\ee
Here we have omitted the index $(i)$.
This form resembles with cascade model \cite{Tanimoto} though our mass matrix is
universal including the quark sector.
For $s\neq 0$, $\hat{M}$ has a small correction as given by,
\be
\hat{M}=\hat{M}(s=0)+\Delta \hat{M}.
\ee
Here, the small correction part $\Delta \hat{M}$ is given by
\be
\Delta \hat{M}=s\frac{2}{\sqrt{6}}(-3m_3+m_2+2m_1)\left\{\mbox{cos}\delta\left(
\begin{array}{ccc}
0&1&-1\\
1&r&0\\
-1&0&-r
\end{array}
\right)
+i\mbox{sin}\delta\left(
\begin{array}{ccc}
0&-1&1\\
1&0&r\\
-1&-r&0
\end{array}
\right)\right\},
\ee
where
\be
r=\frac{2(m_2-m_1)}{-3m_3+m_2+2m_1}.
\ee
It is interesting that the breaking pattern of the 2-3 symmetry in the universal mass matrix has a specific structure. \\

In conclusion, we have presented a simple relation between the CKM and the MNS mixing matrices, 
which is derived from a hypothesis of universal mixing for up and down quarks and neutrinos. 
From the observed constraints from the CKM matrix and the $\theta_{12}$ and $\theta_{23}$ in the MNS mixing matrix, 
we have predicted the magnitude of the mixing angle $\theta_{13}$  
and the CP violating Dirac phase $\delta $ in the MNS matrix. 
Implication of the hypothesis to the mass matrix is also discussed. 
If the predictions derived from the hypothesis are correct, it will give a guiding principle for model building.

\section*{acknowledgement}
The work of T.F.\ is supported in part by the Grant-in-Aid for Science Research
from the Ministry of Education, Science and Culture of Japan
(No.~020540282).

\newpage
{\scalebox{0.6}{\includegraphics{fig1.eps}} }
\begin{quotation}
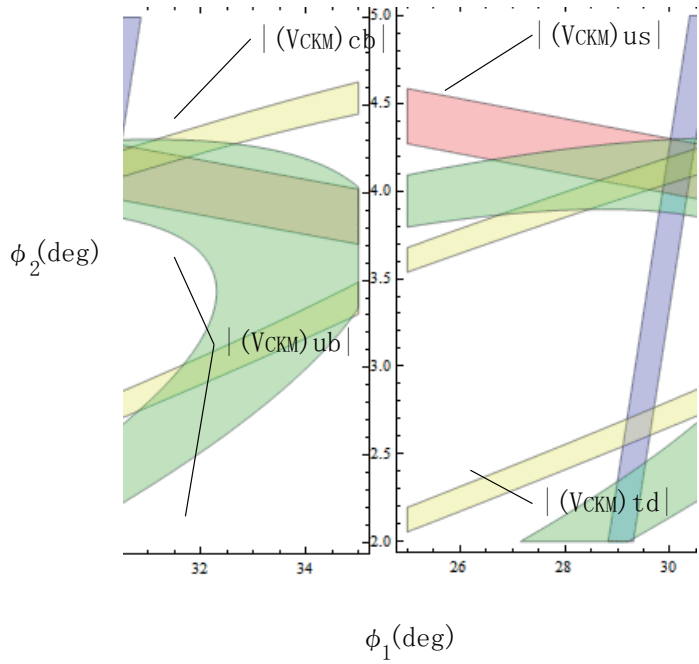
{\bf Fig.~1}  Allowed region in the $\phi_1-\phi_2$ plane 
which is consistent with the experimental data of the CKM quark mixing matrix elements 
for a typical set of the MNS lepton parameters $s_{13}=0.042$ and $\delta=9^\circ $ for an example. 
The shaded areas are allowed, which are obtained from 
the experimental data of $|(V_{CKM})_{us}|$, $|(V_{CKM})_{cb}|$, $|(V_{CKM})_{ub}|$, and $|(V_{CKM})_{td}|$.  
\end{quotation}

\vspace{5mm}

{\scalebox{0.6}{\includegraphics{fig2.eps}} }
\begin{quotation}
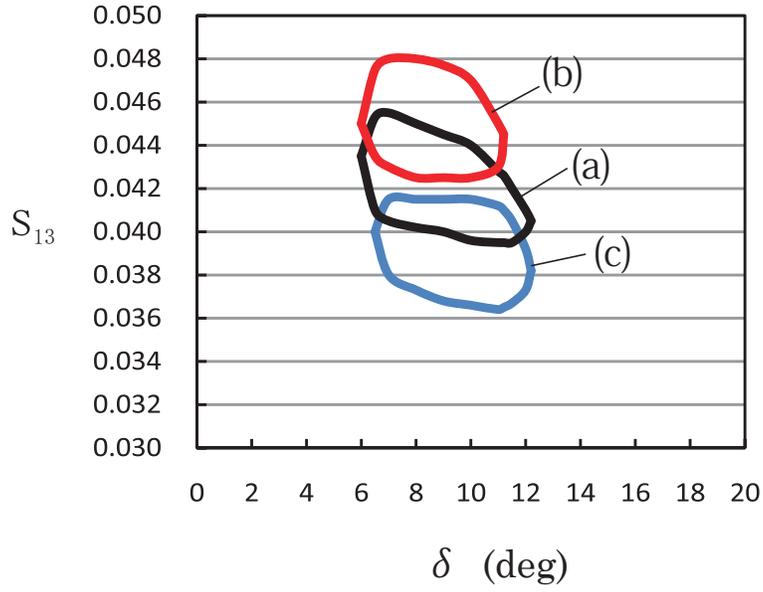
{\bf Fig.~2}  Allowed regions in the $\delta-s_{13}$ plane 
which can reproduce the experimental data of the CKM quark mixing matrix elements.
The inside areas of the solid curves are allowed for the case 
(a) $\mbox{tan}^2\theta_{12}=0.47$ and  $\mbox{sin}^2 2\theta_{23}=1$, 
(b) $\mbox{tan}^2\theta_{12}=0.53$ and $\mbox{sin}^2 2\theta_{23}=1$, 
and (c) $\mbox{tan}^2\theta_{12}=0.42$ and $\mbox{sin}^2 2\theta_{23}=1$.
\end{quotation}


\begin{thebibliography}{99}

\bibitem{F-N}
T. Fukuyama and H. Nishiura, Proceeding of 1997 Shizuoka Workshop on Masses and Mixings of Quarks and Leptons
 (World Scientific, Singapore, 1998), p252; hep-ph/9702253.
\bibitem{Fuku2}
K. Matsuda, T. Fukuyama, and H. Nishiura, Phys. Rev. {\bf D61}, 053001 (2000).
\bibitem{KamLAND}
K. Eguchi et al., [KamLAND Collaboration], Phys. Rev. Lett. {\bf 90}, 021802 (2003).
\bibitem{KamLAND2}
S. Abe et al., [KamLAND Collaboration], Phys. Rev. Lett. {\bf 100}, 221803 (2008).
\bibitem{MINOS}
D.G. Michael et al., [MINOS Collaboration], Phys. Rev. Lett. {\bf 97}, 191801 (2006);
J. Hosaka et al., [Super-Kamiokande Collaboration], Phys. Rev. {\bf D74}, 032002 (2006); 
\bibitem{TBM}
P.F. Harrison, D.H. Perkins, and W.G. Scott, Phys. Rev. Lett. {\bf B530}, 167 (2002).
\bibitem{Koide}
Y. Koide, H. Nishiura, K. Matsuda, T. Kikuchi, and T. Fukuyama, Phys. Rev. {\bf D66}, 093006 (2002).
\bibitem{F-O}
T. Fukuyama and N. Okada, JHEP {\bf 0211}, 011 (2002).
\bibitem{Antusch}
S.Antusch,J.Kersten,M.Lindner,and M,Ratz, Nucl.Phys.{\bf B674}, 401 (2003)
\bibitem{Mohapatra}
Though their proposition of universality is quite different from ours, such degenerate case was discussed by\\ 
S.K. Agarwalla, M.K. Parida, R.N. Mohapatra, and G. Rajasekaran, Phys. Rev. {\bf D75}, 033007 (2007).

\bibitem{PDG}
C. Amsler et al., Particle Data Group, Phys. Lett. {\bf B667}, 1 (2008).
\bibitem{Fogli}
G. Fogli, E. Lisi, A. Marrone, A. Palazzo, and A.M. Rotunno, arXiv:0905.3549[hep-ph].
\bibitem{Tanimoto}
N. Haba, R. Takahashi, M. Tanimoto, K. Yoshioka, Phys. Rev. {\bf D78}, 113002 (2008). 


\end{thebibliography}
\end{document}